\documentclass[12pt,preprint]{aastex}

\newcommand{\etal}{\it{et al.}\rm}
\newcommand{\hst}{{\it Hubble} Space Telescope \rm}
\newcommand{\hsts}{{\it Hubble} Space Telescope's \rm}
\newcommand{\sst}{{\it Spitzer} Space Telescope \rm}
\newcommand{\ssts}{{\it Spitzer} Space Telescope's \rm}

\slugcomment{Accepted by The Astrophysical Journal}

\shorttitle{NML Cyg and the Cyg OB2 Association}
\shortauthors{Schuster et al.}

\begin{document}

\title{Imaging the Cool Hypergiant NML Cygni's Dusty \\ Circumstellar Envelope with Adaptive Optics}

\author{M. T. Schuster, M. Marengo, J. L. Hora, G. G. Fazio,}
\affil{Harvard-Smithsonian Center for Astrophysics, Cambridge, MA 02138}
\email{mschuster@cfa.harvard.edu, mmarengo@cfa.harvard.edu}

\author{R. M. Humphreys, R. D. Gehrz,}
\affil{University of Minnesota, School of Physics and Astronomy, Minneapolis, MN 55455}

\and

\author{P. M. Hinz, M. A. Kenworthy, W. F. Hoffmann}
\affil{Steward Observatory, University of Arizona, Tucson, AZ 85721}

\begin{abstract}
  We present sub-arcsec angular resolution, high-Strehl ratio mid-IR
  adaptive optics images of the powerful OH/IR source and cool
  hypergiant NML Cyg at 8.8, 9.8 and 11.7 $\mu$m. These images reveal
  once more the complexity in the dusty envelope surrounding this
  star. We spatially resolve the physical structures (radius $\sim$
  0.\arcsec14, $\sim 240$ AU adopting a distance of 1.74~kpc)
  responsible for NML Cyg's deep 10 $\mu$m silicate dust absorption
  feature. We also detect an asymmetric excess, at separations of
  $\sim$0.\arcsec3 to 0.\arcsec5 ($\sim$520 to 870 AU), NW from the
  star. The colors of this excess are consistent with thermal emission
  of hot, optically thin dust. This excess is oriented in the
  direction of the Cyg OB2 stellar association, and is likely due to
  the disruption of NML Cyg's dusty wind with the near-UV radiation
  flux from the massive hot stars within Cyg OB2. This interaction was
  predicted in our previous paper \citet{schuster06a}, to explain the
  geometry of an inverted photo-dissociation region observed at
  optical wavelengths. 
\end{abstract}




\section{Introduction}

The powerful OH/IR source and cool hypergiant NML Cyg is one of the
most massive and luminous M stars in the Galaxy \citep[$\sim$40
M$_{\odot}$, $M_{bol} \simeq -9.5$,][]{hyland72, humphreys79,
  morris83, humphreys94, schuster06a, schuster07}. Originally discovered
by \citet{neugebauer65}, NML Cyg was easily identified as an extremely
bright infrared source ($I \sim 8$, $K \sim 1$ mag). The star's relatively
close proximity to the Cyg OB2 stellar association confirms its high
luminosity \citep{morris83,schuster06a}. Even though NML Cyg is luminous,
it is also heavily obscured at visual wavelengths due to high interstellar
(IS) and circumstellar (CS) extinction, with $V$ fainter than 16.6 mag.
Its visual/near--IR spectrum indicates a M6 spectral type \citep{wing67},
and its spectrum peaks in the 2--20 $\mu$m range
\citep{gillett68,stein69}. Thus, the spectrum indicates a substantial,
optically thick dusty CS envelope obscuring the central star. NML Cyg's
thick envelope is a result of its strong post-Main Sequence (post-MS) wind
(with a velocity of 23 km~s$^{-1}$)  and high continuous mass-loss rate of
$6.4\times10^{-5}$ M$_{\odot}$ yr$^{-1}$ \citep{hyland72, bowers81,
morris83}. \citet{habing82} mapped an unusual oblong shaped HII
region\footnote{See references therein for the history of the discovery.}
at 21 cm nearly 1{\arcmin} in size around NML Cyg, revealing another
unique feature of this enigmatic M supergiant. In addition, more recent
high-angular resolution, high-contrast images from the \hst ({\it HST})
WFPC2 camera reveal an asymmetric CS nebula of dust scattered stellar
light \citep{schuster06a}. The CS nebula has a shape similar to the HII
region, but is about 300 times smaller in radius. It is necessary to
consider NML Cyg's local environment and its effects on the star in order
to understand this object's true nature.

Although NML Cyg is a very luminous post-MS star which is losing mass at a
prodigious rate, it does not currently dominate its local interstellar
environment because of its unique location within the Milky Way. It lies
in relatively close proximity to Cyg OB2, which is possibly the largest OB
stellar association -- in size, mass and number of OB stars -- in the
Galaxy \citep[see][and references therein for a multi-wavelength
review]{knodlseder03}. The Cyg OB2 association spans nearly 2$^{\circ}$ on
the sky, or $\sim$30 pc in radius at the distance of $1.74\pm0.2$ kpc
\citep{massey91}, making it one of the closest massive associations to the
Sun. Cyg OB2 has $\sim$120 O star members, including 5 of the 10 most
luminous O stars in the Galaxy \citep{humphreys78}, and approximately 2500
B stars. The stellar mass within the association is possibly as high as
10$^{5}$ M$_{\odot}$. The O stars provide the UV radiation responsible for
a Str\"{o}mgren sphere that extends to a radius of at least 2.$^{\circ}$74
on the sky, or $\gtrsim$80 pc (the projected separation between Cyg OB2
and NML Cyg). The Lyman continuum photon flux of the whole association is
estimated to be 10$^{51}$ ph~s$^{-1}$, or $\gtrsim$10$^{9}$
ph~cm$^{-2}$s$^{-1}$ at the location of NML Cyg. Mid--IR images from the
Midcourse Space eXperiment (MSX) satellite show NML Cyg is located in a
relative void of gas/dust inside the Cygnus X super bubble
\citep{schuster07}. This configuration allows the Lyman continuum photons
and near--UV radiation from the hot stars within Cyg OB2 to travel the
large distance to NML Cyg relatively unimpeded.

\citet{morris83} demonstrated that the Lyman continuum radiation from
Cyg OB2 is responsible for the asymmetric HII region observed around
NML Cyg. The HII region is the result of a photo-ionization
interaction between the spherically symmetric, expanding hydrogen wind
from NML Cyg balanced against an incident plane parallel Lyman
continuum photon flux from Cyg OB2 (see Figures 1 and 2 in
\citealt{morris83}). The photo-ionization is inverted in the sense
that usually the nearest massive star is the ionization source, and
the ionization occurs outward from the central
source. \citeauthor{morris83} also demonstrated that the strength of
the Lyman continuum flux from Cyg OB2 and the density of atomic
hydrogen around NML Cyg are sufficient to produce the observed 21 cm
emission, thus confirming NML Cyg's high continuous mass-loss rate. In
a similar way, \citet{schuster06a} described how near--UV photons with
energies between a few and 13.6 eV photo-dissociate the molecular gas
in NML Cyg's wind. \citeauthor{schuster06a}'s model explains why the
asymmetric shape of the CS nebula resolved in \hst WFPC2 images is a
consequence of this physical interaction with Cyg OB2. Recent
observations of excited-state OH maser emission in NML Cyg's wind by
\citet{sjouwerman07} are consistent with an increased OH column
density, presumably arising from dissociated H$_{2}$O molecules at the
water photo-dissociation region. Since the atomic and molecular gas
around NML Cyg is disrupted, the dust facing Cyg OB2 is unprotected
from the near-UV photons. It is thus likely that the CS dust grains
are heated and destroyed by the radiation from Cyg OB2 and should emit
a detectable infrared signature.

In Section \ref{obs}, we present sub-arcsec angular resolution, high
Strehl Ratio MIRAC3/BLINC mid-IR images of NML Cyg obtained at the 6.5
meter MMT observatory with the Adaptive Optics (AO) system. These
observations are among the highest resolution images of this star at
mid-IR wavelengths, and reveal the complexity in the asymmetric dust
envelope surrounding NML Cyg. New photometry from the \ssts Infrared
Array Camera, the \hsts WFPC2 camera, and ground-based infrared
observations are presented in Section \ref{obs}. The Spectral Energy
Distribution (SED) for NML Cyg's CS envelope presented in Section
\ref{sed} shows an optically thick silicate dust absorption
feature. In Section \ref{aoimages}, we characterize the resolved
structures in the dusty CS envelope. In Section \ref{discuss}, we show
that the CS dust component facing Cyg OB2 exhibits a silicate feature
in emission. We suggest that UV radiation from the hot, massive stars
within Cyg OB2 likely causes the inversion in the silicate feature
through external heating and dust destruction. We conclude by
discussing these results in relation to previous works on NML Cyg, and
comparing NML Cyg to the extremely luminous M--type hypergiant VY CMa.

\section{Observations}\label{obs}
\subsection{MMT/AO Observations}

\begin{figure}
\begin{center}
\includegraphics[width=0.95\textwidth,angle=0]{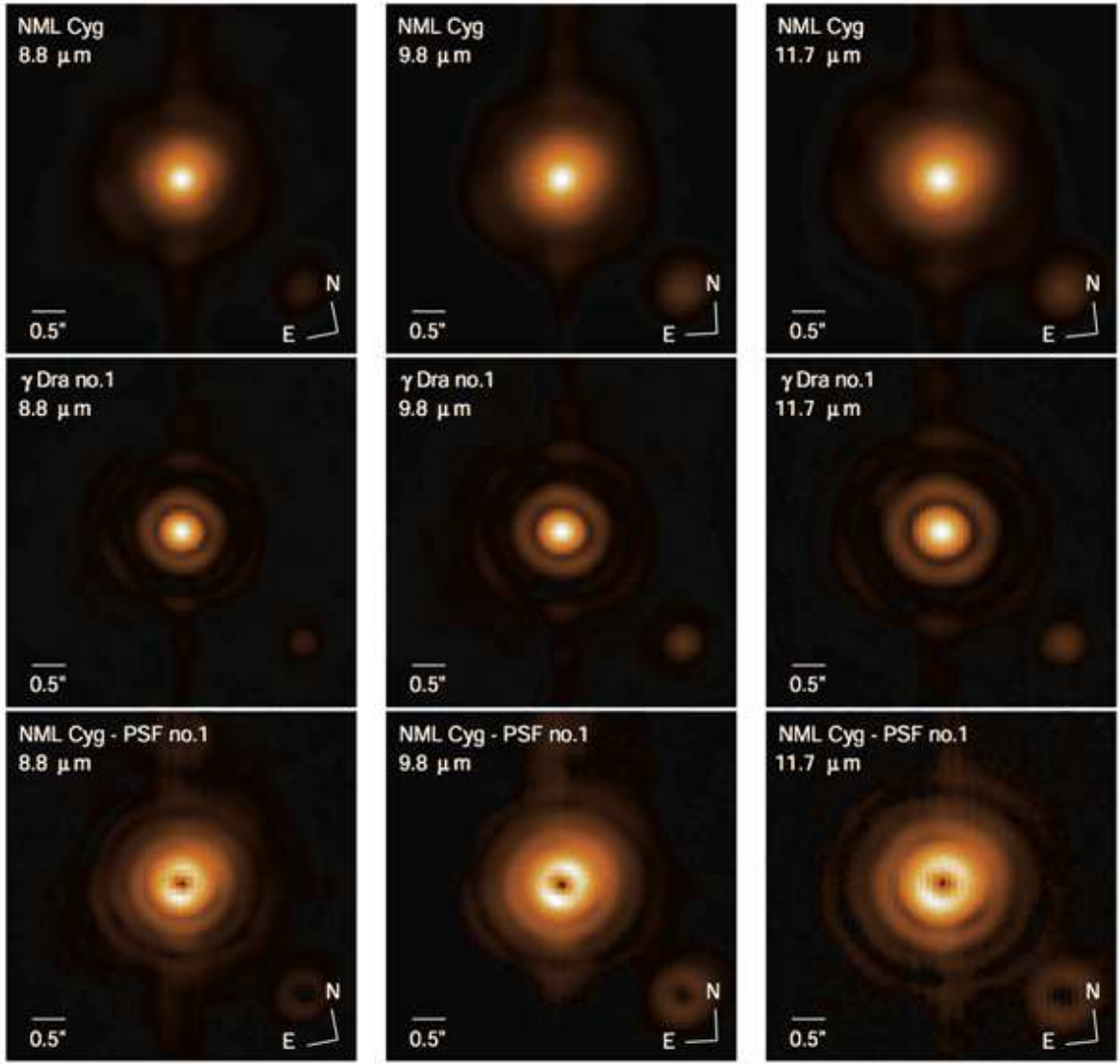}
\end{center}
\caption{ \label{fig1} 8.8, 9.8 and 11.7 $\mu$m MMT/AO MIRAC3/BLINC
  images of NML Cyg's dusty CS nebula. \textit{Top row --} Images of
  NML Cyg, no post-processing. \textit{Middle row --} PSF reference
  star images ($\gamma$ Dra, epoch 1). \textit{Bottom row --} NML Cyg
  after over-subtracting the PSF images above (scaled to peak
  brightness to give zero flux). NML Cyg's CS envelope is clearly
  broader than a point-source, extended with a NW/SE
  orientation. There is also an asymmetric excess on the NW side. The
  spot in the lower right corner is a PSF ghost. The images are
  displayed with a square root intensity scale.}
\end{figure}

The inverse Photo-Dissociation Region (PDR) model for NML Cyg's CS
envelope described in \citet{schuster06a} implies the presence of warm
dust near the photo-dissociation front(s). According to this model,
Cyg OB2's UV radiation is heating/destroying the CS dust on the side
facing Cyg OB2. In order to corroborate this model through direct
observation, we obtained high-angular resolution, high-contrast
mid--IR images of NML Cyg's dusty CS nebula. The observations were
made using the University of Arizona's adaptive optics system
\citep{wildi03} and the MIRAC3/BLINC camera \citep{hoffmann98, hinz00}
on the 6.5 meter MMT telescope. Table \ref{tab1} summarizes the
observations made in July 2006 and Figure \ref{fig1} shows the NML Cyg
images at 8.8, 9.8 and 11.7 $\mu$m. The BLINC module was used in
`chop' mode with a frequency of $\sim$1 Hz, which is useful for
background (sky) subtraction when combined with `nod' dithers. Using
natural guide star mode, the MMT/AO system produced stable,
high-Strehl ratio ($\sim$$96-98$\%), high signal-to-noise, nearly
diffraction limited Point-Spread Functions (PSF) at 10 $\mu$m. We have
used the source itself as guide star for the AO system (while extended
in the infrared, NML Cyg is a point source at optical wavelengths
where the AO system operates). With a magnitude $R \sim 14$, NML Cyg
is at the sensitivity limit for the MMT/AO wavefront sensor, but was
successfully used thanks to the excellent sky conditions.

Figure \ref{fig1} also shows images of the IR standard $\gamma$ Dra,
which was observed at two separate epochs and sky orientations for
calibration and PSF characterization (see Table \ref{tab1}, only one
epoch is shown in Figure \ref{fig1}). The PSF stability was excellent
throughout the night and the sky conditions were especially good, with
seeing $\la$0.\arcsec25. In these PSF images the minima/maxima in the
diffraction pattern are closely matched to the Airy pattern for a
6.35m aperture\footnote{The MMT adaptive secondary is undersized to
  improve IR performance, resulting in a 6.35m equivalent resolution
  limit.}, with a FWHM of 0.\arcsec33 at 8.8 $\mu$m, 0.\arcsec36 at
9.8 $\mu$m, and 0.\arcsec44 at 11.7 $\mu$m.

The nearly diffraction limited images have similar benefits in
resolution and stability to images from space-based observatories, and
at mid--IR wavelengths we can detect and separate the externally
heated dust from the rest of NML Cyg's CS nebula. Relatively long
exposure times and a 5 point dither pattern, repeated multiple times
with sub-pixel shifts, provided very high dynamic range images of NML
Cyg and the calibration/PSF standards, in the range $\sim 2 \times
10^{3}-10^{4}$ (PSF peak relative to background noise). The images
were processed with the Drizzle algorithm to maximize the quality of
the data and recover sampling resolution \citep{hook97, koekemoer00,
  fruchter02}. The processed MIRAC3 images were resampled by
2$\times$2 from 0.\arcsec0954 pix$^{-1}$ to a scale of 0.\arcsec0477
pix$^{-1}$.

The 8.8, 9.8 and 11.7 $\mu$m filters were chosen to provide good
spectral coverage of the 10 $\mu$m silicate dust feature observed for
this luminous cool star \citep{monnier97, blocker01}. The photometry
of the source in the three MIRAC3 bands is listed in
Table~\ref{tab2}. The flux in each band was measured in an aperture of
2\arcsec{} radius, with the sky residual background flux (after sky
subtraction for each chop/nod sequence) determined outside this
aperture. Photometric calibration was obtained using the PSF reference
star $\gamma$~Dra as photometric standard. The photometry has been
corrected to a common airmass. The uncertainty in the photometry
quoted in Table~\ref{tab2} do not include 5 -- 10\% systematic
uncertainties typical for ground based mid-IR observations.

\subsection{{\it Spitzer}/IRAC Observations}

Table \ref{tab1} summarizes our observations of NML Cyg obtained in
July 2004 with the Infrared Array Camera \citep[IRAC,][]{fazio04}
on-board the \sst \citep{werner04,gehrz07}. The IRAC images (GTO
Program 124, AOR 6588416, pipeline version S14.0.0) were processed
with IRACproc, a software package that facilitates co-addition and
analysis of IRAC data \citep{schuster06b}. IRACproc also improves the
identification and removal of cosmic rays and other transient signals
from the processed images, while simultaneously improving image
quality, photometric accuracy and sensitivity (signal-to-noise). Due
to the extreme brightness of the source, the core of the IRAC images
was saturated, preventing the adoption of standard aperture photometry
techniques to measure the flux in the IRAC bands. By fitting a High
Dynamic Range (HDR) PSF to the diffraction spikes and extended
``wings'' of the IRAC images, we have however derived the source
photometry in the IRAC bands at 3.6, 4.5 and 8.0~$\mu$m (artifacts in
the 5.8~$\mu$m image prevented using this technique at that
wavelength). The PSFs we used were obtained from the observation of
bright stars (Vega, $\epsilon$~Eridani, Fomalhaut and $\epsilon$~Indi)
as part of the \sst GTO program 90, and are available at the
\emph{Spitzer} Science Center web
site\footnote{http://ssc.spitzer.caltech.edu/irac/psf.html}. The
detailed description of this fitting procedure is available in
\citet{marengo06} and \citet{schuster06b}. The photometry is listed in
Table \ref{tab2}.

\subsection{Ground-Based IR Photometry}

We obtained ground-based broadband 1 to 12 $\mu$m photometry in August
2000 at the University of Minnesota's (UMN) O'Brien Observatory (OBO)
with a single element Bolometer using a standard chop-nod technique
for background sky subtraction \citep[beam diameter $\sim$
27\arcsec,][]{gehrz74, gehrz92, gehrz97}. The observations and
photometry are summarized in Tables \ref{tab1} and \ref{tab2}. The
absolute calibration of the instrument was determined using
$\alpha$~Lyr (Vega) and $\beta$~Peg as primary calibrator, using the
procedure described in \citet{gehrz97}.

\subsection{{\it HST}/WFPC2 PC Photometry}

Multi-wavelength images of NML Cyg were obtained in September 1999
with the WFPC2 Planetary Camera on-board the \hst
\citep{biretta00}. These observations aimed to investigate the star's
CS nebulosity are summarized in Table \ref{tab1} \citep[images
previously published by][]{schuster06a}. NML Cyg was observed with
broadband Johnson-Cousins $V$ and $R$ filters as well as narrow band
H$_{\alpha}$ filter. Prior to co-addition, the images were processed
with the Space Telescope Science Institute's standard calibration
reference files. The images were combined with the IRAF/STSDAS
software package DITHER, which uses the Drizzle algorithm to recover
image resolution from the pixel response of the camera while
preserving photometric accuracy
\citep{koekemoer00,fruchter02}. Multiple, dithered exposures allowed
removal of cosmic rays, bad pixels, and other effects during
co-addition. The {\it HST} Photometry is reported in Table \ref{tab2}.

\section{NML Cyg's Bolometric Luminosity and \\ 10 $\mu$m Silicate Dust Absorption Feature}\label{sed}

\begin{figure}
\epsscale{.80}
\includegraphics[width=0.75\textwidth,angle=-90]{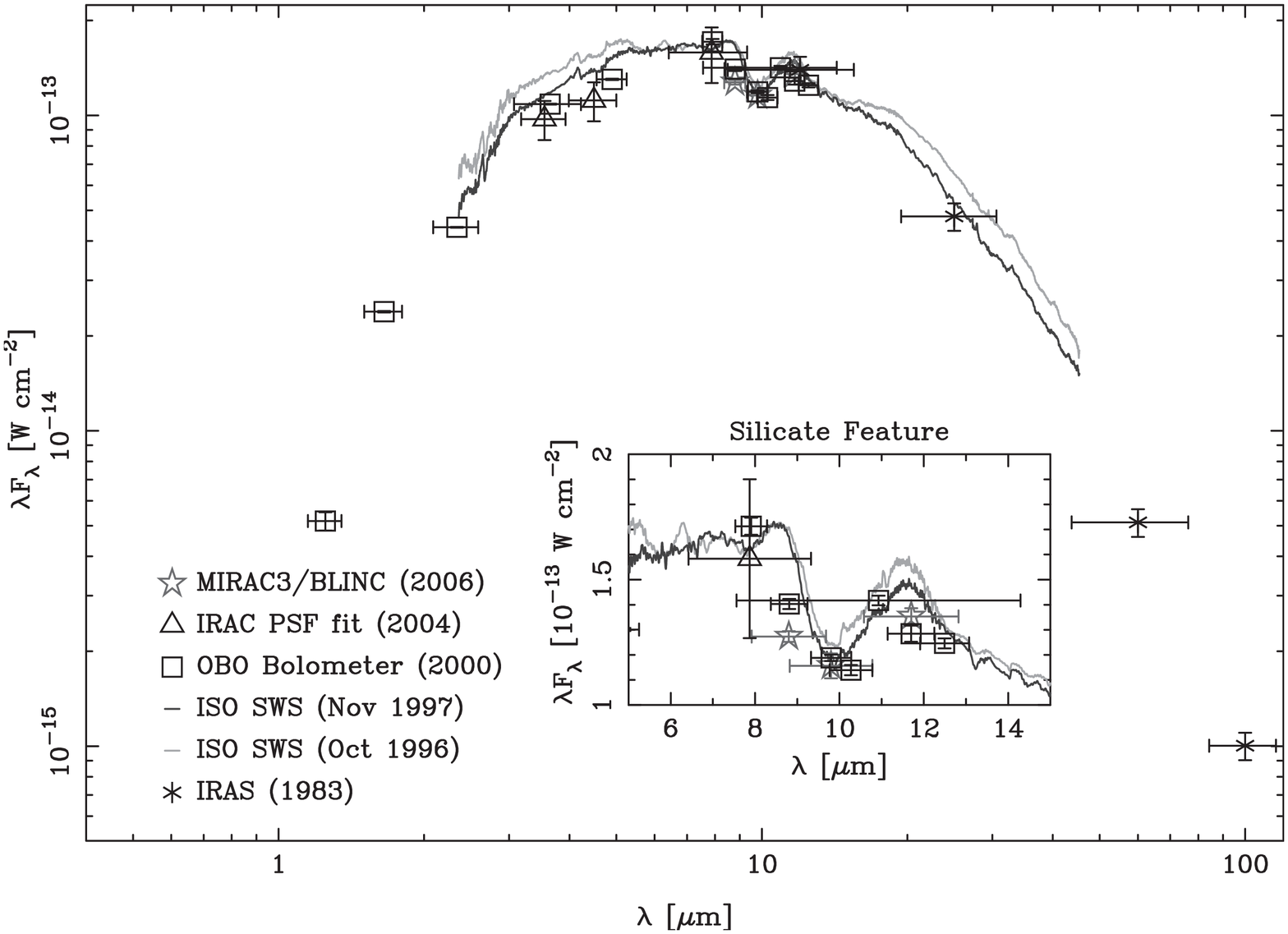}
\caption{ \label{fig2} NML Cyg SED from photometry in Table
  \ref{tab2}, as well as ISO SWS spectra
  \citep{justtanont96,kraemer02}. The 10 $\mu$m silicate feature is
  seen in absorption, and is a result of the optically thick CS
  envelope enshrouding the star. There is some normalization offset
  between the ISO spectra and the other observations. At least some of
  this difference is presumably from the variation in the star's
  spectrum over its 3 yr period. The ground-based observational errors
  have typical $5-10$\% systematic uncertainties (only calibration
  uncertainties are plotted here). NML Cyg's extreme brightness led to
  large IRAC PSF fitting errors. The star's high luminosity may have
  led to saturation in the IRAS observations, resulting in the offset
  with the ISO spectra.}
\end{figure}

Table \ref{tab2} summarizes our photometric observations of NML Cyg
along with archival IRAS Point Source fluxes. These observations are
plotted on the ${\lambda}F_{\lambda}$ vs. $\lambda$ SED in Figure
\ref{fig2}, along with reprocessed archival ISO SWS spectra
\citep{justtanont96,kraemer02}.

NML Cyg's spectrum rises rapidly from optical wavelengths through the
IR to peak in the 2 to 20 $\mu$m range, with a broad far--IR
tail. Visible light images presented by \citet{schuster06a} show that
NML Cyg is almost completely obscured by the CS envelope enshrouding
the star, thus confirming significant circumstellar extinction. This
is evident from the broad 10 $\mu$m silicate dust absorption
feature. Due to the optically thick envelope, this feature is a
striking characteristic of the NML Cyg spectrum, setting this star
apart from other cool hypergiants such as $\mu$ Cep \citep{gehrz70},
S Per \citep[][and unpublished data]{humphreys74}, VY CMa
\citep{smith01}, VX Sgr \citep{humphreys74}, HR 5171 A
\citep{humphreys71}, IRC +10420 \citep{jones93} and M33 Var A
\citep{humphreys06}, where the 10 $\mu$m silicate feature is seen in
emission from optically thin CS nebula shells.

NML Cyg is also a semi-regular variable with period $\sim 940$ days
\citep[][and references therein]{monnier97}. The vertical spread
between data sets in Figure \ref{fig2} is well within the $\sim 50$\%
amplitude variation reported by \citet{monnier97} from 8 to 13
$\mu$m. Even with this large variability, \citeauthor{monnier97}'s
results show the 10 $\mu$m silicate feature's shape is roughly
constant over a span of nearly 19 years. In addition, the vertical
offsets between observations at other wavelengths in Figure \ref{fig2}
are at least in part due to variability. However, the degree to which
NML Cyg's spectrum varies, in either amplitude or shape, at other
wavelengths has not been well established. Interstellar (IS)
extinction to NML Cyg is also high, with an estimated range of $A_{V}
\sim 3.7-4.6$ magnitudes \citep{lee70, gregory76}.

We calculate NML Cyg's minimum bolometric luminosity by
integrating\footnote{Using a Rayleigh-Jeans $F_\lambda \propto
  \lambda^{-4}$ fit extension at long wavelengths.} the SED in Figure
\ref{fig2}, obtaining L$_{bol}$ = $1.04\pm0.05 \times 10^{5} \cdot
(d/kpc)^{2}$ L$_{\odot}$, or $3.15\pm0.74$ $\times$ 10$^{5}$
L$_{\odot}$ at $1.74 \pm 0.2$ kpc \citep[][Cyg OB2 distance modulus
11.2]{massey91}. This luminosity does not include a correction for IS
extinction, which would add at most a few percent to the total since
most of the light is in the mid--IR where extinction is low. The
uncertainty in this estimate does not consider the changes in the
source brightness due to its variability. This luminosity value is in
good agreement with the estimate of L$_{bol}$ =
$1.13\times10^{5}$$\cdot$($d$/kpc)$^{2}$ L$_{\odot}$ by
\citet{blocker01}. Thus, NML Cyg's intrinsic luminosity, a few
$\times$~10$^{5}$~L$_\odot$ ($M_{bol}$ $\simeq$ -9.0) is similar to
that of other extremely luminous M--type hypergiants.

Previously, \citet{morris83} reported a luminosity of
$\sim5\times10^{5}$ L$_{\odot}$ for NML Cyg at a distance of 2.0
kpc. This luminosity was equal to the most luminous known M
supergiants in the Milky Way and LMC, $M_{bol} \simeq -9.5$,
$\sim5\times10^{5}$L$_{\odot}$ \citep{humphreys78, humphreys79}. Our
value is 37\% lower, with 24\% due to the downwardly revised distance
and 13\% from our integration of the SED from 0.5 to 100 $\mu$m
(including new photometry). With this revision, NML Cyg appears to be
less luminous than the hypergiant VY CMa (M4-5e Ia, $4.3\times10^{5}$
L$_{\odot}$ at 1.5 kpc, also without correcting for IS extinction),
which has a comparable SED \citep[][]{smith01, humphreys07}. The
uncertainty in our estimate, and the variability of the source makes
it difficult to determine unambiguously which of these stars is more
luminous. Even with this reduction in luminosity, however, NML Cyg
remains the most luminous star with spectral type M6 or later known in
the Milky Way.

\section{Mid--IR Image Analysis}\label{aoimages}

Figure \ref{fig1} shows that NML Cyg's CS envelope is clearly extended
at mid--IR wavelengths, compared to the point-source (PSF) images of
$\gamma$ Dra, as the diffraction pattern appears `filled-in'. As
further evidence for the extension and degree of asymmetry, Figure
\ref{fig1} also shows the NML Cyg images after over-subtracting the
PSF (epoch 1, scaled to give zero intensity flux at the center pixel
in the subtracted images). The asymmetric CS envelope residuals are
apparent, having a NW/SE elongation axis, and position angle in the
range $\sim120^{\circ}-150^{\circ}$. Also visible in the
PSF-subtracted images is a lopsided tail extending to the NW (across
the first diffraction ring). The MMT is an Elevation/Azimuth mount
telescope, so the background sky predictably rotates in the image
plane while tracking an object. Thus, it was possible to directly
confirm the asymmetric extension by observing NML Cyg twice in the
same night, at two sky rotation angles separated by
$\sim$122$^{\circ}$ (at 9.8 $\mu$m only). The extended components
rotated by the same number of degrees (not shown here), eliminating
the possibility that the PSF or other instrumental effects are the
cause of the observed extension.

NML Cyg's mid--IR envelope orientation is similar to the
$130^{\circ}-150^{\circ}$ position angles observed in the OH
\citep{masheder74, benson79, diamond84}, H$_{2}$O \citep{richards96},
and SiO \citep[and private communication]{boboltz04}
masers. \citet{richards96} have suggested that the NW/SE spatial
distribution of the H$_{2}$O vapor masers may indicate a bipolar
outflow, supported by recent SiO ground state observations by
\citeauthor{boboltz04}. It is also possible that the maser emission is
tracing an asymmetric, episodic outflow that may be reminiscent of the
arcs and other structures seen in the circumstellar nebula surrounding
VY CMa \citep{smith01,smith04,humphreys05,humphreys07,jones07}.

Earlier observations by \citet{hinz01} also show NML Cyg's extended
envelope using MIRAC3 with the BLINC module in nulling interferometer
mode on the MMT. These observations were made prior to the
installation of the adaptive optics system. \citeauthor{hinz01}'s
observations indicated a CS envelope extended beyond 0.\arcsec28 at
10.3 $\mu$m, with nearly equal flux in the extended envelope relative
to the unresolved structure (see his Eq. 7.5 and Table 7.2,
p. 89-95). The better resolution, and Fourier $u,v$ coverage, of the
AO images in Figure \ref{fig1} reveal at least two structures in the
CS nebula, the main envelope aligned with maser observations and an
asymmetric component aligned more closely with the direction towards
Cyg OB2.

\subsection{Parameterizing the CS Envelope Image Intensity}

\begin{figure}
\includegraphics[width=0.3\textwidth,angle=-90]{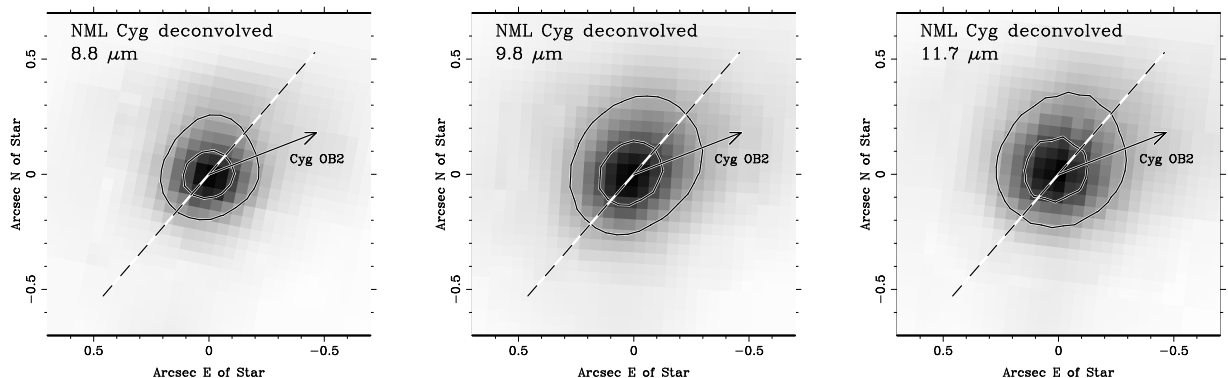}
\caption{ \label{fig3} MMT/AO MIRAC/BLINC images of NML Cyg after
  deconvolution with the PSF reference star ($\gamma$~Dra, epoch
  1). The inner contour is set to the source 1/2 maximum flux level,
  and shows that the core of the emission is elongated along the SE --
  NW axis ($\sim 140^{\circ}$, dashed line). The outer contour is set
  to 1/10 of the source maximum flux level, and shows that the
  external part of the circumstellar emission is lopsided towards the
  NW quadrant. The images are displayed with a square root intensity
  scale. The arrow indicates the direction of the Cyg OB2
  association.}
\end{figure}

To take full advantage of the exceptional stability and contrast in
the MMT/AO MIRAC3 high-angular resolution images, we used
deconvolution to more clearly reveal details seen in the Figure
\ref{fig1} images of NML Cyg's CS envelope. Figure \ref{fig3} shows
our NML Cyg images after deconvolution with the PSF calibrator
\citep[using IRAF task LUCY,][]{richardson72, lucy74, snyder90}. The
deconvolution reveals the complex CS envelope structure, highlighting
the elongated envelope extended towards the NW
direction. The two contours in the Figure are set at the source 1/2
and 1/10 maximum flux level. The inner contour shows that the core of
the emission is elongated along a SE -- NW axis, while the external
part of the circumstellar emission is lopsided towards the NW
quadrant. The main axis of the core emission has a PA of $\sim
140^\circ$ (counter-clockwise from north) in the 9.8 and 11.7~$\mu$m
deconvolved images. This angle is similar to the alignment of the NML
Cyg OH, H$_2$O and SiO masers, also at $\sim 130^\circ$ --
150$^\circ$. The PA is slightly smaller ($\sim 120^\circ$) in the
8.8~$\mu$m image, possibly due to an image artifact in direction of
the MIRAC3/BLINC chop.

\begin{figure}
\epsscale{.60}
\plotone{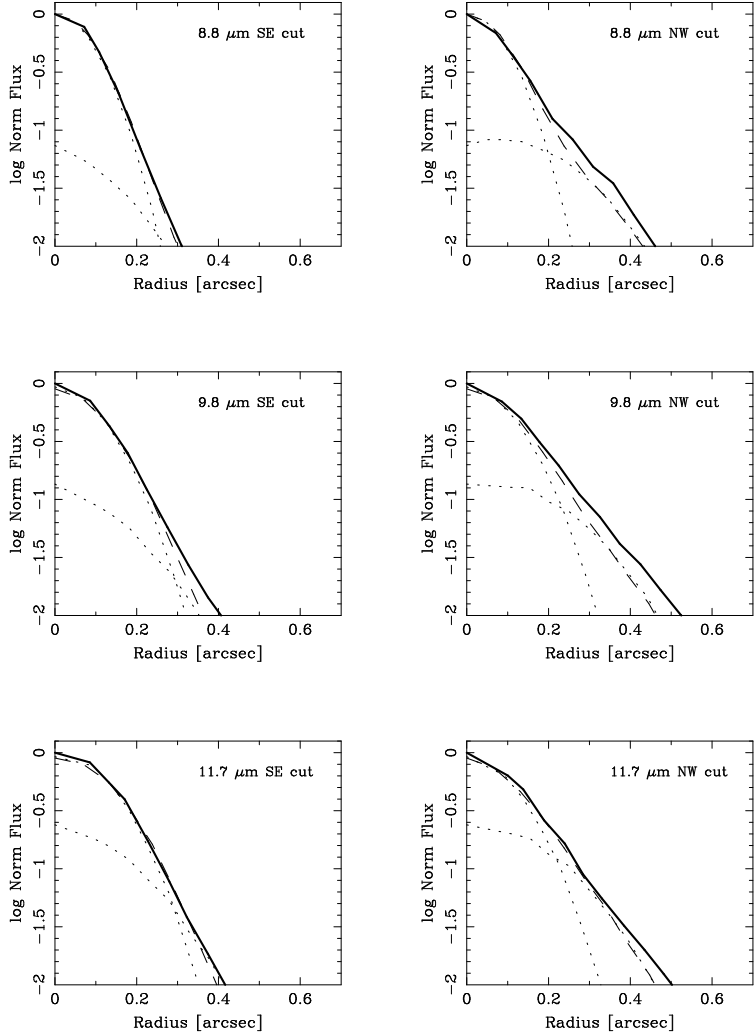}
\caption{ \label{fig4} \textit{Solid line} --- Radial profile of the
  NML Cyg deconvolved images in the SE (left) and NW (right)
  directions. The profile is derived along the dashed line plotted in
  Figure~\ref{fig3}, averaged on a 60$^\circ$ wedge. \textit{Dashed
    line} --- Radial profile of the 2-Gaussian fit in the same
  directions. \textit{Dotted lines} --- Radial profile of the
  individual Gaussian best fit components. Note how the Gaussian fit
  describes well the profile in the SE direction, but a significant
  excess is left in the NW direction. The deconvolved image profiles
  are normalized to the peak intensities, and the Gaussian profiles
  have been rescaled accordingly.}
\end{figure}

To better characterize the shape and spatial scales of the NML Cyg
circumstellar envelope, we fit the images in Figure~\ref{fig3} with two
2-D Gaussians, one fitting the core emission (which we call the ``core
envelope'' Gaussian) and one fitting the NW extended low level structures
(the ``outer envelope'' Gaussian). Note that this fitting process is aimed
only to find a practical way of describing the changing colors of the
circumstellar envelope at different distances from the center. The fitting
process or results do not imply that the individual Gaussians represent
individual physical structures in the circumstellar envelope. The core
Gaussian has a width and orientation analogous to the width and position
angle of the 1/2 height contour, and is centered on the star. The second
Gaussian fits the outer envelope emission, and is offset in the NW
direction as shown by the 1/10 contour level asymmetry. The best fit
parameters were determined by minimizing the residuals after subtraction
of the two Gaussians from the deconvolved images. We start by fitting the
first Gaussian to the image core, then the second Gaussian is fit to the
core-subtracted image. The process was then iterated to minimize the
residuals after subtraction of both Gaussians. Table~\ref{tab3} lists the
best fit parameters for each wavelength, including the size parameter
$\sigma$ and relative flux (in \% of the total flux) of each component,
the aspect ratio $a/b$ between the major and minor semi-axis and the
position angle PA of the core component and the offset of the outer
envelope with respect to the core. Figure~\ref{fig4} shows the radial
profiles of the deconvolved image and best fit Gaussians along the cut at
PA = 140$^\circ$. Note how the profiles show the large asymmetry between
the SE and NW directions. While the SE profile is fit well by the two
Gaussians, some residual flux above the fit is still present in the NW
direction.

The agreement in position angle and size between the core envelope
emission and the NML Cyg maser observations suggest that this emission
arises from the NML Cyg's wind at, or just interior to, the
photo-dissociation front(s) described in \citet{schuster06a}. The
relative flux contribution from the outer envelope component increases
from 8.8 to 9.8 to 11.7~$\mu$m as we are resolving the optically thin
warm dust ($T_{dust} \sim 200$ -- 500~K) farther out in the envelope.

\subsection{The Asymmetric NW Excess}
 
\begin{figure}
\includegraphics[width=0.3\textwidth,angle=-90]{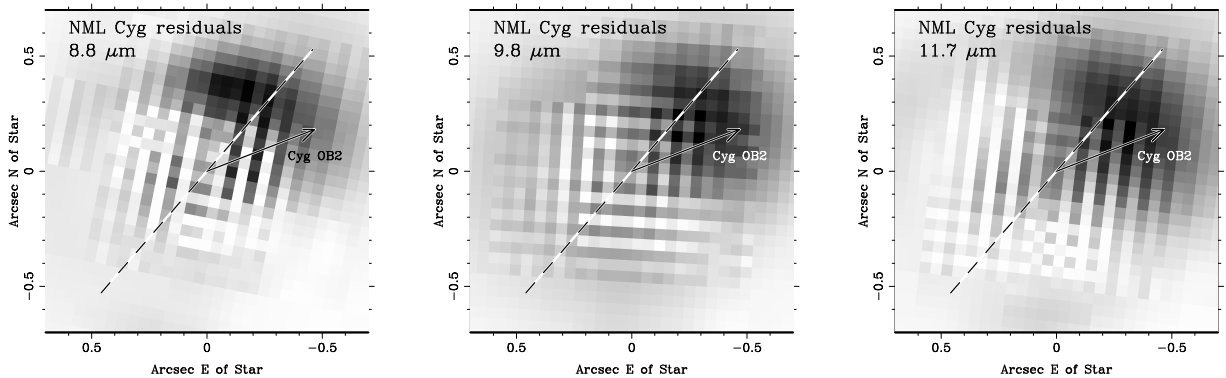}
\caption{ \label{fig5} Residuals after subtraction of the 2-Gaussian
  fit, convolved with the PSF reference ($\gamma$ Dra, 1st epoch),
  from the NML Cyg MMT/AO MIRAC3/BLINC images. The residuals show a
  clear asymmetric excess in the NW quadrant, roughly in the direction
  of the Cyg OB2 association. This asymmetric excess also agrees with
  the lopsidedness seen in the {\it HST}/WFPC2 images from
  \cite{schuster06a}. External heating by the UV radiation from Cyg
  OB2 likely causes this excess emission. The bright spot to the North
  in the 8.8 $\mu$m image is possibly bleeding from MIRAC3's vertical
  chop. Grid-like patterned residuals represent the limiting noise due
  to the image spatial sampling on the coarse detector array grid.}
\end{figure}

To test the accuracy of the deconvolution parametrization in image
space, we have subtracted the Gaussian fit, convolved with the PSF
reference ($\gamma$ Dra), from the NML Cyg mid-IR images at each
wavelengths. The residuals are shown in Figure~\ref{fig5}. The
grid-like pattern is due to the spatial sampling and linear
interpolation of the image on the detector grid, and is the limiting
factor in the accuracy of the fit. This pattern may be reduced by
increasing the number of unique offsets in the dithered observations
and by applying higher order interpolation in the reconstruction of
the image on a finer pixel grid. The uniform distribution of this
residual pattern and the relatively low flux level in the center is
indicative of the good quality of the deconvolutions and fits, and
also the stability of the MMT/AO PSF.

The figure also shows a clear excess residual in the NW quadrant. The
excess peak brightness occurs around $\sim 0$.\arcsec3 to 0.\arcsec5
($\sim520-870$ AU) from the central star, with a broad tail extending
toward Cyg OB2. This excess cannot be reduced by adjusting the
position and shape of the outer envelope Gaussian, constrained by the
fit in the other 3 quadrants. The shape and orientation of this excess
are reminiscent to the lopsided emission seen in the HST images from
\citet[][their Figures 2 and 6]{schuster06a}. The asymmetric excess is
separated from the star in Figure \ref{fig5} at the spatial resolution
limit, i.e. it is resolved according to the Rayleigh criterion. It
lies just outside the asymmetric dusty reflection nebula seen in the
{\it HST} images \citep{schuster06a}. We suggest that the excess
emission is likely from warm dust that is externally heated/destroyed
by UV radiation from Cyg OB2. In the next section, this hypothesis
will be tested by directly measuring the flux from this excess.

\section{Discussion: CS Dust Heating and Destruction}\label{discuss}

\begin{figure}
\includegraphics[width=0.65\textwidth,angle=-90]{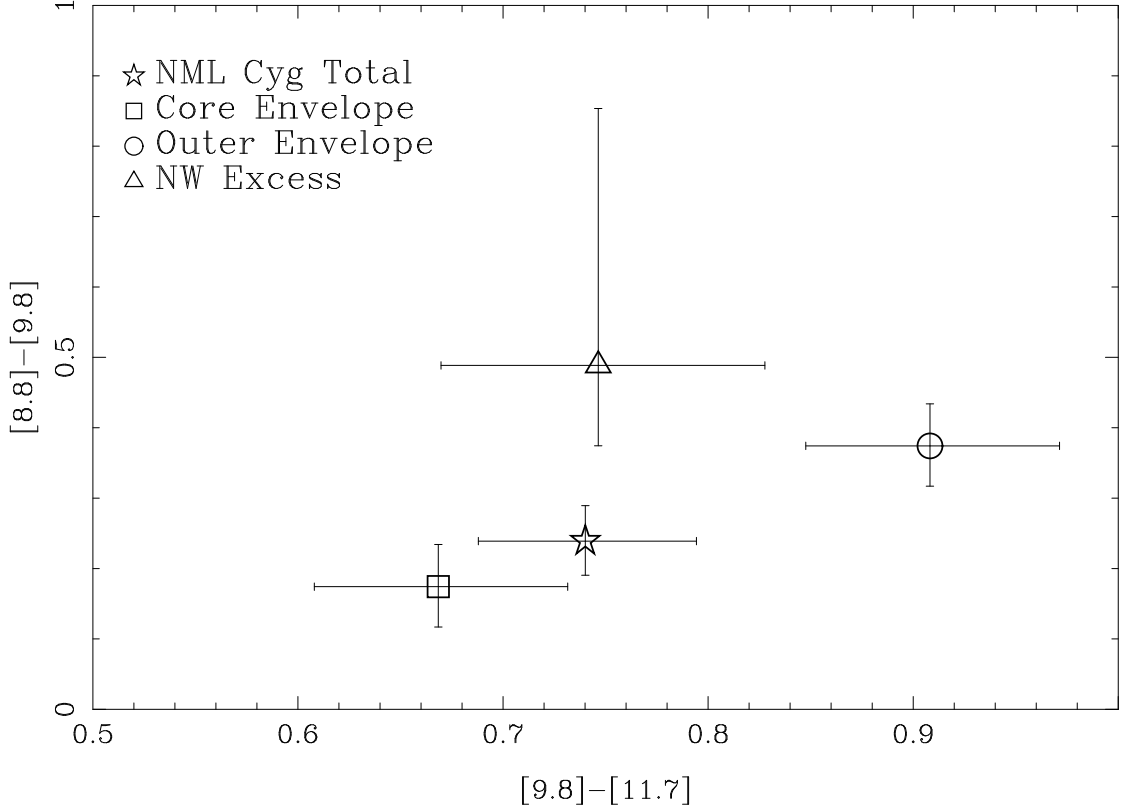}
\caption{ \label{fig6} Mid--IR color-color plot for NML Cyg and its CS
  envelope components from MIRAC3 observations. The envelope's core
  [8.8]-[9.8] and [9.8]-[11.7] colors are blue as compared to the
  total for the source, indicative of hotter and optically thick
  dust. The outer CS envelope component is redder in both colors,
  indicating cooler dust. In contrast, the colors for the asymmetric
  excess are blue in [9.8]-[11.7] and red in [8.8]-[9.8], suggesting
  warm and optically thin (9.8~$\mu$m silicate feature in emission)
  dust. The colors are in the Vega magnitude system. The larger
  uncertainty in the NW excess colors reflects the possible
  contribution from MIRAC3 vertical chop bleeding N of the star at 8.8
  $\mu$m.}
\end{figure}

In Table~\ref{tab4} we list the total flux of the source, the fraction
of the total flux in the core and outer envelope, the total flux of
all the fit residuals shown in Figure~\ref{fig5}, and the excess
residual flux in the NW quadrant (residual flux in the NW quadrant,
minus the residual flux in the opposite SE quadrant). In Figure
\ref{fig6} we plot the [8.8]-[9.8] and [9.8]-[11.7] colors for each CS
nebula component. Both the core envelope colors are bluer than the
total for the source. This core component likely represents the
collective flux from where the silicate absorption is deepest, and
thus characterizes the geometry of the dust responsible for the
optically thick feature seen in the star's spectrum. NML Cyg's stellar
radius is probably of the order of $\sim$10 AU, typical for the
largest red supergiants, and therefore the material comprising the
inner CS envelope ($\sigma\sim0.\arcsec12$, 200 AU) has a size scale
of approximately 20 stellar radii. The outer CS envelope colors are
redder than those of the core envelope and the total source. This
indicates dust with decreasing temperature and density.

In contrast, the colors for the asymmetric excess are significantly
different, being blue in [9.8]-[11.7] and red in [8.8]-[9.8]. These
colors can be explained by the presence of optically thin hot dust
with the 9.8~$\mu$m silicate feature in emission. The fact that the
asymmetric dust emission excess appears only on the side facing Cyg
OB2 is consistent with an increase in temperature resulting from the
external UV radiation. The density is also likely reduced due to dust
destruction. Since the asymmetric excess is shielded from the central
star by the optically thick inner CS envelope, Cyg OB2's external
energy is required to account for changes in the grain temperature,
size and/or spatial distribution ultimately responsible for the
silicate feature's inversion.

In \citet{schuster06a} we showed that the UV flux from the Cyg OB2
association is strong enough to significantly heat the dust grains in
the NML Cyg envelope, possibly destroying the smallest grains. The
excess mid-IR flux we observe in the NW quadrant may be the direct
effect of this heating and destruction. Its presence strongly suggests
that the circumstellar envelope of NML Cyg is indeed shaped by the
interaction of its wind with the radiation field coming from the Cyg
OB2 association. The direct detection of this physical interaction
between NML Cyg and the Cyg OB2 association is further confirmation
that the source is indeed in the neighborhood of Cyg OB2. This
supports the distance determination of $1.74 \pm 0.2$ kpc by
\citet{massey91} for this star as one of the most reliable distances
for an evolved massive star near the upper luminosity boundary.

\section{NML Cyg's Complex Circumstellar Environment}

\begin{figure}
\epsscale{.65}
\plotone{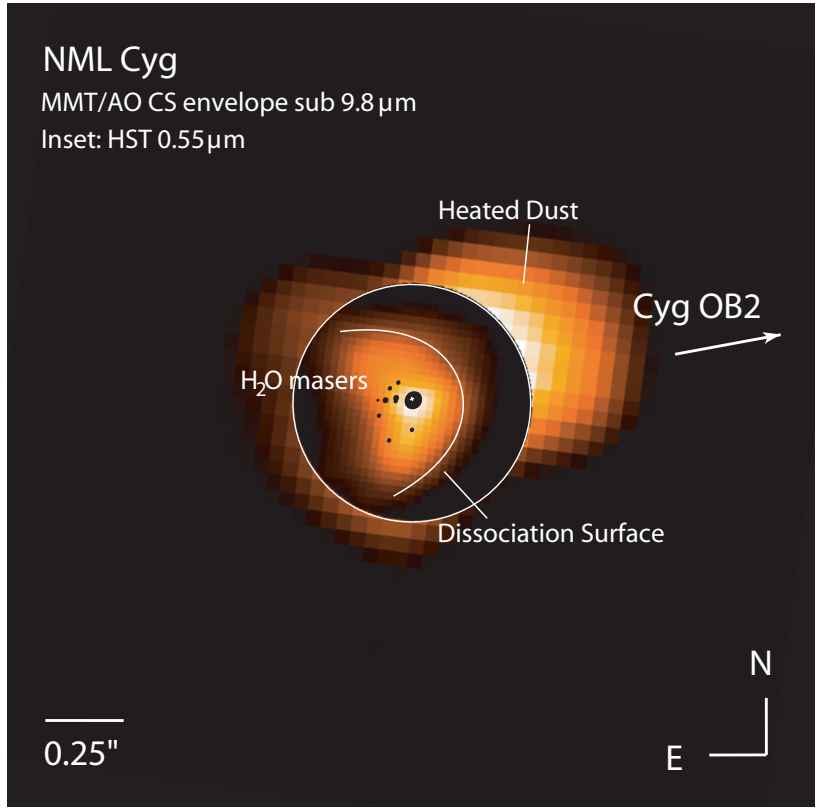}
\caption{ \label{fig7} Composite image of the physical interaction
  taking place in NML Cyg's circumstellar environment. {\it Main
    image} -- the CS dust heated by Cyg OB2 is isolated in this
  deconvolved MMT/AO MIRAC3/BLINC 9.8 $\mu$m image after CS envelope
  model subtraction. {\it Inset} -- this {\it HST}/WFPC2 F555W ($V$
  band, 0.55 $\mu$m) image shows the scattered stellar light
  \citep[reproduced from][]{schuster06a}. The symmetry axis is shown
  in the direction of the nearby Cyg OB2 stellar association. The
  brightest 22 GHz H$_{2}$O maser features from \citet[Figure
  1,][]{richards96}, and the molecular photo-dissociation surface
  projection from \citet[Figure 4,][]{schuster06a} have been
  superimposed on the image. The white mark and large maser near the
  center indicate the star's location, and maser sizes are roughly
  proportional to their flux. For display the dissociation surface is
  placed along the outer edge of the envelope as seen in the {\it HST}
  image. The 9.8 $\mu$m image is displayed with a linear intensity
  scale. The 0.55 $\mu$m image is displayed with a square root
  intensity scale.}
\end{figure}

Figure~\ref{fig7} shows the NW excess in our 9.8~$\mu$m deconvolved
image compared to the scattered light HST/WFPC2 F555W image from
\citet{schuster06a}. As discussed in Section~5, our interpretation for
this mid-IR excess is another strong indicator that the UV radiation
from the hundreds of massive young OB stars within Cyg OB2 is
disrupting NML Cyg's post-MS wind through photo-ionization,
photo-dissociation and grain destruction. The models for these
processes explain the inverted HII region's existence and shape, the
asymmetric nebula seen with {\it HST}, as well as the externally
heated dust visible in our mid-IR images. The inverse photo-ionization
\citep{morris83} and photo-dissociation \citep{schuster06a} models for
this interaction have assumed a radially symmetric, constant velocity
hydrogen gas wind, i.e. density $\rho \propto r^{-2}$. The agreement
between theory and observation suggests that this steady-state,
uniform outflow is a overall a good approximation of NML Cyg's stellar
wind.

However, the asymmetric mid-IR image reveals that NML Cyg's dusty CS
envelope does has a more complex underlying structure. One may
speculate that the identified structures may represent the integrated
signal from many arcs, loops, bipolar outflows and other 3-dimensional
structures below our angular resolution that may be the result of
episodic and asymmetric mass-loss, much like VY CMa (see Figure 3,
\citealt{smith01} as well as \citealt{humphreys07,
  jones07}). Asymmetric structures arising from episodic mass-loss may
also modulate the maser and IR emission as they pass through the
photo-dissociation and grain destruction regions. With even better
angular resolution and sensitivity (perhaps with {\it JWST} and/or the
next generation of ground-based nulling interferometers like LBTI and
KNI) it should be possible to observe these fainter, compact
structures within the larger nebulosity, if they exist.

Previous 11 $\mu$m results from \citet{monnier97}, \citet{danchi01}
and \citet{blocker01} have suggested the presence of multiple,
concentric density enhancements (shells) surrounding the central star
superposed on a $r^{-2}$ wind, characterized by an azimuthally
symmetric, but non-steady-state dust outflow, i.e $\rho_{dust}
\not\propto r^{-2}$. This, in turn, implied a time dependent
mass-loss, possibly in episodic/periodic events such as a `superwind'
phase. The images presented in Section \ref{aoimages} show evidence
for a complex distribution in the CS dust, but also reveal an
asymmetric excess aligned to the NW (towards Cyg OB2). The differences
with these earlier results are not surprising given the better
resolution and Fourier $u,v$ coverage in the MMT/AO images. However,
it should be noted that \citeauthor{monnier97} acknowledge in their
conclusions that deviations from spherical symmetry, particularly
emission from dense or clumped material 350 mas from the central star,
would necessarily change the meaning of their fits \citep[and likewise
for][]{danchi01}. In fact, Figure \ref{fig3} and \ref{fig4} show,
through direct imaging, the presence of warm dust emitting throughout
mid-IR wavelengths precisely at this distance from NML Cyg. Moreover,
this emission is concentrated at position angles ranging from
$\sim-30^{\circ}$ (CS envelope) to $\sim-60^{\circ}$ (dust facing Cyg
OB2), or equivalently at alignments of $\sim$ $120^{\circ}$ --
$150^{\circ}$. \citeauthor{monnier97}'s and \citeauthor{danchi01}'s
journals of observations show that they made eight observations with
resolution substantially (at least 20\%) better than
$\sim$0.\arcsec35, and these observations were made with the
interferometer aligned along position angles between 102$^{\circ}$ and
140$^{\circ}$ (see each author's Table 1). Thus, their observations
just happened to be oriented with the NW excess emission resolved in
our images. It is possible that the most external circumstellar shell
inferred from these interferometric observations may in fact be better
explained by the asymmetric excess in our 9.8 $\mu$m image.

Using a combination of Keck aperture masking and IOTA interferometry,
\citet{monnier04} found at 2.2~$\mu$m an elongated structure with a
diameter of $\sim 50$~mas. The orientation of this structure is
orthogonal to the PA of the inner component of our images, represented
by our ``core'' Gaussian fits. \citet{monnier04} interpret this
structure as an equatorial enhancement in the dust envelope,
orthogonal to NML Cyg's maser outflow.

Our mid-IR images, when analyzed in combination with the previous
observations discussed above, clearly demonstrate the complexity of
NML Cyg circumstellar environment. Any model of NML Cyg that attempts
to explain the observations must thus take into account the observed
asymmetries as well as the physical interaction between NML Cyg's wind
and the UV radiation from Cyg OB2. In particular for NML Cyg,
high-angular resolution ($\la 0$.\arcsec3) combined with more complete
coverage of the Fourier $u,v$ plane are crucial to further investigate
asymmetries in its complex CS environment.

NML Cyg is part of our larger program to study the circumstellar
environments of cool stars near the empirical upper luminosity boundary in
the HR diagram with evidence for high-mass loss and observed
instabilities. As such, NML Cyg was perhaps the best candidate to have an
extensive CS nebula, possibly like VY CMa. However, NML Cyg's CS
nebulosity is more concentrated around the star and appears quite
different compared to VY CMa, as imaged by \citet{smith01}. Even though
NML Cyg is heavily obscured, the extent of the CS nebula is much less than
the almost 10{\arcsec} nebula surrounding VY CMa ($\ga 10^{4}$ AU at VY
CMa's distance of 1.5 kpc). It is possible that the more distant material
in NML Cyg's circumstellar nebula has been largely dissipated by the winds
and radiation pressure inside the Cygnus-X super bubble and is below our
detection limits.  If NML Cyg were not located in such close proximity to
Cyg OB2, it might show a much more extended nebula comparable to VY CMa.
However, VY CMa's mass-loss rate, $2-4\times10^{-4}$ M$_{\odot}$yr$^{-1}$
\citep{danchi94}, is 3 to 6 times higher than for NML Cyg and this
difference may also have significant bearing on the more extensive
circumstellar nebulosity.

\section{Conclusions}

Our sub-arcsec angular resolution AO mid-IR images of NML Cyg provide
a new understanding of the complex geometry and physics of the
circumstellar environment of this high luminosity cool hypergiant. By
spatially resolving the optically thick dusty envelope, we directly
image the structures responsible for the creation of a deep 10 $\mu$m
silicate absorption feature. This structure, which follows the same
orientation of NML Cyg's maser outflow, is orthogonal to a near-IR
equatorial enhancement found by \citet{monnier04}.

By analyzing the mid-IR colors of structures located at increasing
distance from the star, we observe a trend in which the optical depth
of the dust decreases in the outer parts of the circumstellar
envelope. For the first time we isolate an asymmetric excess, at a
distance of $\sim 520$ to 870 AU, NW from the star, with colors
consistent with the emission of hot, optically thin dust. We interpret
this emission as the signature of the interaction of the NML Cyg
circumstellar envelope with the strong UV flux generated in the nearby
Cyg OB2 association. This interaction was predicted in our previous
paper \citep{schuster06a} to explain the shape of a the inverted
photo-dissociation region we imaged with the HST at optical
wavelength. Our new mid-IR observations strongly support the validity
of our previous results and the model we proposed for their
explanation.

\acknowledgments

We are pleased to acknowledge interesting conversations with David
Boboltz regarding his VLA observations. NASA provided support for this
work. Observations reported here were obtained at the MMT Observatory,
a joint facility of the Smithsonian Institution and the University of
Arizona. This work is based in part on observations made with the {\it
  Spitzer} Space Telescope, which is operated by the Jet Propulsion
Laboratory, California Institute of Technology under a contract with
NASA. Support for RDG was provided by NASA through grants 1215746 and
1256406 issued by JPL-Caltech to the University of Minnesota. This
work is based in part on observations made with the NASA/ESA {\it
  Hubble} Space Telescope, obtained at the Space Telescope Science
Institute, which is operated by the Association of Universities for
Research in Astronomy, Inc., under NASA contract NAS5-26555.

{\it Facilities:} \facility{MMT}, \facility{Spitzer}, \facility{HST}, \facility{UMN OBO}.



\clearpage






\begin{deluxetable}{llllr}
\tabletypesize{\scriptsize}
\renewcommand{\arraystretch}{0.7}
\tablewidth{0pt}
\tablecaption{Observations\label{tab1}}
\tablehead{MMT AO MIRAC3 & $\lambda$ & $\Delta\lambda$ & Airmass & Exposures \\
July 23, 2006 UT & ($\mu$m) & ($\mu$m) & & no.$\times$sec}
\startdata
NML Cyg & 8.80 & 0.88 & 1.10 & 13$\times$20 \\
. & 9.80 & 0.98 & 1.18 & 11$\times$20 \\
. & 9.80 & 0.98 & 1.02 & 2$\times$20 \\
. & \hspace{-0.5em}11.70 & 1.12 & 1.14 & 11$\times$20 \\
\multicolumn{5}{l}{} \\
$\gamma$ Dra (calibrator) & 8.80 & 0.88 & 1.08 & 15$\times$30 \\
. & 8.80 & 0.88 & 1.18 & 15$\times$30 \\
. & 9.80 & 0.98 & 1.07 & 20$\times$30 \\
. & 9.80 & 0.98 & 1.23 & 36$\times$30 \\
. & \hspace{-0.5em}11.70 & 1.12 & 1.09 & 20$\times$30 \\
. & \hspace{-0.5em}11.70 & 1.12 & 1.28 & 20$\times$30 \\[0.2em]
\tableline \\[-0.5em]
{\it Spitzer}/IRAC & $\lambda$ & $\Delta\lambda$ & AOR & Exposures \\
July 27, 2004 UT & ($\mu$m) & ($\mu$m) & & no.$\times$sec \\[0.2em]
\tableline \\[-0.5em]
NML Cyg & 3.550 & 0.750 & 6588416 & 3$\times$10.4\tablenotemark{a} \\
. & 4.493 & 1.015 & . & 3$\times$10.4\tablenotemark{a} \\
. & 5.731 & 1.425 & . & 3$\times$10.4\tablenotemark{a} \\
. & 7.872 & 2.905 & . & 3$\times$10.4\tablenotemark{a} \\[0.2em]
\tableline \\[-0.5em]
OBO MN Bolometer & $\lambda$ & $\Delta\lambda$ & Filter & Exposure \\
Aug 10, 2000 UT & ($\mu$m) & ($\mu$m) & & \\[0.2em]
\tableline \\[-0.5em]
NML Cyg & 1.250 & 0.200 & J & n/a\\
. & 1.653 & 0.297 & H & . \\
. & 2.340 & 0.500 & K & . \\
. & 3.647 & 1.152 & L & . \\
. & 4.900 & 0.700 & M & . \\
. & \hspace{-0.5em}10.925 & 6.730 & N & . \\
. & 7.908 & 0.755 & & . \\
. & 8.808 & 0.871 & & . \\
. & 9.803 & 0.953 & & . \\
. & \hspace{-0.5em}10.273 & 1.013 & & . \\
. & \hspace{-0.5em}11.696 & 1.110 & & . \\
. & \hspace{-0.5em}12.492 & 1.157 & & . \\[0.2em]
\tableline \\[-0.5em]
{\it HST}/WFPC2 & $\lambda$ & $\Delta\lambda$ & Filter & Exposures \\
Sept 16, 1999 UT & ($\mu$m) & ($\mu$m) & & no.$\times$sec \\[0.2em]
\tableline \\[-0.5em]
NML Cyg & \hspace{-0.5em}0.4293 & 0.0473 & F439W (B) & 6$\times$500\tablenotemark{b} \\
. & \hspace{-0.5em}0.5337 & 0.1228 & F555W (V) & 20, 100, 4$\times$400 \\
. & \hspace{-0.5em}0.6564 & 0.0022 & F656N (H$_{\alpha}$) & 20, 2$\times$260 \\
. & \hspace{-0.5em}0.6677 & 0.0867 & F675W (R) & 0.5, 10 \\
\enddata
\tablecomments{Exposure time not applicable to UMN Bolometer since the instrument measures changes in Voltage to obtain instrumental magnitudes and signal-to-noise.}
\tablenotetext{a}{ Saturated.}
\tablenotetext{b}{ No detection.}
\end{deluxetable}

\clearpage






\begin{deluxetable}{llllrr}
\tabletypesize{\scriptsize}
\renewcommand{\arraystretch}{0.7}
\tablewidth{0pt}
\tablecaption{NML Cyg Photometry\label{tab2}}
\tablehead{
\colhead{Instrument} & \colhead{\hspace{6em}} & \colhead{Filter} & \colhead{$\lambda$ ($\mu$m)} & \colhead{Flux (Jy)} & \colhead{$\sigma_{F}$ (Jy)}
}
\startdata
\multicolumn{6}{l}{{\it HST}/WFPC2} \\
\multicolumn{6}{l}{Sept 16, 1999 UT} \\[0.2em]
\tableline \\[-0.5em]
 & & V & 0.5337 & 2.377e-04 & 0.003e-04 \\
 & & H$_{\alpha}$ & 0.6564 & 2.395e-03 & 0.014e-03 \\
 & & R & 0.6677 & 6.124e-03 & 0.029e-03 \\[0.2em]
\tableline \\[-0.5em]
\multicolumn{6}{l}{OBO MN Bolometer\tablenotemark{a}} \\
\multicolumn{6}{l}{Aug 10, 2000 UT} \\[0.2em]
\tableline \\[-0.5em]
 & & J & 1.250 & 21.5 & 1.4 \\
 & & H & 1.653 & 131 & 0.7 \\
 & & K & 2.340 & 333 & 1.0 \\
 & & L & 3.647 & 1287 & 4.2 \\
 & & M & 4.900 & 2121 & 10 \\
 & & N & \hspace{-0.5em}10.925 & 4325 & 58 \\
 & & & 7.908 & 4522 & 90 \\
 & & & 8.808 & 4124 & 60 \\
 & & & 9.803 & 3881 & 43 \\
 & & & \hspace{-0.5em}10.273 & 3905 & 71 \\
 & & & \hspace{-0.5em}11.696 & 5020 & 120 \\
 & & & \hspace{-0.5em}12.492 & 5366 & 84 \\[0.2em]
\tableline \\[-0.5em]
\multicolumn{6}{l}{{\it Spitzer}/IRAC PSF fitting\tablenotemark{b}} \\
\multicolumn{6}{l}{July 27, 2004 UT} \\[0.2em]
\tableline \\[-0.5em]
 & & 3.6 & 3.550 & 1150 & 160 \\
 & & 4.5 & 4.493 & 1670 & 240 \\
 & & 8.0 & 7.872 & 4160 & 830 \\[0.2em]
\tableline \\[-0.5em]
\multicolumn{6}{l}{MMT AO MIRAC3/BLINC\tablenotemark{a}} \\
\multicolumn{6}{l}{July 23, 2006 UT} \\[0.2em]
\tableline \\[-0.5em]
 & & & 8.80 & 3735 & 63 \\
 & & & 9.80 & 3780 & 160 \\
 & & & \hspace{-0.5em}11.70 & 5280 & 130 \\[0.2em]
\tableline \\[-0.5em]
\multicolumn{6}{l}{IRAS PSC\tablenotemark{c}} \\
\multicolumn{6}{l}{1983} \\[0.2em]
\tableline \\[-0.5em]
 & & & \hspace{-0.5em}12 & 5580 & 560 \\
 & & & \hspace{-0.5em}25 & 3990 & 400 \\
 & & & \hspace{-0.5em}60 & 1030 & 100 \\
 & & & \hspace{-1.0em}100 & 335 & 34 \\
\enddata
\tablecomments{Photometry not color corrected.}
\tablenotetext{a}{ Observational errors exclude typical $5-10$\% systematic uncertainties.}
\tablenotetext{b}{ NML Cyg's mag is near the limit for IRAC PSF fitting \citep[see][]{schuster06b}. 5.8 $\mu$m PSF fitting was not reliable.}
\tablenotetext{c}{ Fluxes from the IRAS Point Source Catalog rejects -- from the InfraRed Science Archive: http://irsa.ipac.caltach.edu, data tag: ADS/IRSA.Gator{\#}2007/1009/122926{\_}27251. 10\% flux uncertainty assumed.}
\end{deluxetable}

\clearpage






\begin{deluxetable}{llrrrrll}
\tabletypesize{\scriptsize}
\rotate
\renewcommand{\arraystretch}{0.7}
\tablewidth{0pt}
\tablecaption{CS Envelope Component Parameters\label{tab3}}
\tablehead{
\colhead{$\lambda$ ($\mu$m)} & \colhead{Comp.} & \colhead{$\Delta$RA
  (mas) \tablenotemark{a}\tablenotemark{b}} & 
\colhead{$\Delta$Dec (mas)\tablenotemark{a}\tablenotemark{b}} &
\colhead{Flux (\%)\tablenotemark{c}} & 
\colhead{$\sigma$ (mas)\tablenotemark{d}\tablenotemark{e}} &
\colhead{$a/b$} & \colhead{P.A. ($^{\circ}$)\tablenotemark{f}}
}
\startdata
 8.8 $\mu$m & & & & & & & \\
            & core envelope  &   0 &   0 & 74.1 &  91.0 & 1.09 & 120 \\
            & outer envelope &  14 &  87 & 25.9 & 171.7 & 1.00 & --- \\
 9.8 $\mu$m & & & & & & & \\
            & core envelope  &   0 &   0 & 70.4 & 124.0 & 1.20 & 138 \\
            & outer envelope &  44 & 105 & 29.6 & 176.5 & 1.00 & --- \\
11.7 $\mu$m & & & & & & & \\
            & core envelope  &   0 &   0 & 65.6 & 113.1 & 1.02 & 139 \\
            & outer envelope &  45 &  98 & 34.4 & 171.7 & 1.00 & --- \\
\enddata
\tablenotetext{a}{ Position offsets are relative to core
  component, +$\Delta$RA is to the West} 
\tablenotetext{b}{ The errors, including fit uncertainty and field
  rotation during the observation, is $\sim 5$~mas.}
\tablenotetext{c}{ Percentages are relative to total
  flux. Uncertainties better than 2\%.} 
\tablenotetext{d}{ FWHM = 2$\sqrt{2\ln{2}}$ $\times$ $\sigma$}
\tablenotetext{e}{ Uncertainty of 10~mas or better}
\tablenotetext{f}{ Major axis ($a$) position angle is measured in
  degrees counter-clockwise from North} 
\end{deluxetable}

\clearpage






\begin{deluxetable}{llr@{\hspace{0.3em}}lr@{\hspace{0.3em}}l}
\tabletypesize{\scriptsize}
\renewcommand{\arraystretch}{0.7}
\tablewidth{0pt}
\tablecaption{CS Envelope Component Fluxes\label{tab4}}
\tablehead{
\colhead{$\lambda$ ($\mu$m)} & Comp. & \multicolumn{2}{l}{Flux (Jy)\tablenotemark{a}}}
\tablecolumns{4}
\startdata
8.8 $\mu$m & & & & \\
& Total Flux     & 3735 & $\pm$ 63 \\
& Core Envelope  & 2540 & $\pm$ 67 \\
& Outer Envelope &  888 & $\pm$ 23 \\
& Total Residual &  307 & \\
& NW$-$SE Excess \tablenotemark{b} &   95 & \hspace{0.3em}$_{-27}^{+10}$ \\
& Vega & 49.5    &      & \\[0.3em]
9.8 $\mu$m & & & & \\
& Total Flux     & 3780 & $\pm$ 160 \\
& Core Envelope  & 2422 & $\pm$ 114 \\
& Outer Envelope & 1018 & $\pm$  48 \\
& Total Residual &  343 & \\
& NW$-$SE Excess &  121 & $\pm$ 4 \\
& Vega & 40.2    &      & \\[0.3em]
11.7 $\mu$m & & & & \\
& Total Flux     & 5280 & $\pm$ 130 \\
& Core Envelope  & 3167 & $\pm$ 101 \\
& Outer Envelope & 1660 & $\pm$  53 \\
& Total Residual &  453 & \\
& NW$-$SE Excess &  170 & $\pm$ 11 \\
& Vega & 28.4    &      & \\[0.3em]
\enddata
\tablecomments{Photometry not color corrected.}
\tablenotetext{a}{ Observational errors exclude typical $5-10$\% systematic 
uncertainties.}
\tablenotetext{b}{ Negative error includes uncertainties in
  subtracting the  vertical chop bleeding.} 
\end{deluxetable}

\clearpage

\clearpage

\clearpage

\clearpage

\clearpage

\clearpage

\end{document}